TITLE PAGE

# Synthesis, crystal structure, microstructure, transport and magnetic properties of SmFeAsO and SmFeAs(O$_{0.93}$F$_{0.07}$)

A. Martinelli[1,*], M. Ferretti[1,2], P. Manfrinetti[1,2], A. Palenzona[1,2]
M. Tropeano[1,3], M. R. Cimberle[4], C. Ferdeghini[1], R. Valle[3], M. Putti[1,3], A.S. Siri[1,3]

[1] *CNR-INFM-LAMIA Artificial and Innovative Materials Laboratory, Corso Perrone 24, 16152 Genova – Italy*
[2] *Dept. of Chemistry and Industrial Chemistry, University of Genoa, via Dodecaneso 31, 16146 Genova – Italy*
[3] *Dept. of Physics, University of Genoa, via Dodecaneso 33, 16146 Genova – Italy*
[4] *CNR-IMEM via Dodecaneso 33, 16146 Genova – Italy*

* corresponding author: amartin@chimica.unige.it

**Abstract**
SmFeAsO and the isostructural superconducting SmFeAs($O_{0.93}F_{0.07}$) samples were prepared. Characterization by means of Rietveld refinement of X-ray powder diffraction data, scanning electron microscope observation, transmission electron microscope analysis, resistivity and magnetization measurements were carried out. Sintering treatment strongly improves the grain connectivity, but, on the other hand, induces a competition between the thermodynamic stability of the oxy-pnictide and $Sm_2O_3$, hence worsening the purity of the sample. In the pristine sample both magnetization and resistivity measurements clearly indicate that two different sources of magnetism are present: the former related to Fe ordering at $T \sim 140$ K and the latter due to the Sm ions that orders antiferromagnetically at low temperature. The feature at 140 K disappears in the F-substituted sample and, at low temperatures a superconducting transition appears. The magnetoresistivity curves of the F-substituted sample probably indicates very high critical field values.

## 1. Introduction

Recently the class of compounds referred to as Fe-based oxy-pnictides, *RE*FeAsO (*RE* = rare earth), attracted much attention on account of the recent discovery of superconductivity in F-substituted LaFeAsO [1] at $T_c = 26$ K. These phases crystallize in the tetragonal system and their structure is built up by planar layers of edge sharing tetrahedra stacked perpendicular to the *c*-axis. Two kinds of layers are present, characterized by different chemical compositions: in the former case the tetrahedra are centred by O with the *RE* at vertices (charge reservoir layer), in the latter one the tetrahedra are centred by Fe with As at vertices (conducting layer). The layers are then linked by As-*RE* bonds. This crystal structure is strongly related to that of $RE_2CuO_4$, a class of materials investigated in the past for their superconducting properties [2,3]; both cuprates and Fe-pnictides share the same kind of tetrahedral *RE*-bearing layer, but in the former case the conducting layer is constituted of vertices sharing squares centred by Cu. A further difference is that in *RE*FeAsO the *RE* centres a slightly distorted square anti-prism whereas in $RE_2CuO_4$ it occurs in tetragonal prism coordination.

Pristine compounds presents a feature at around 140-150 K which has been observed with a lot of experimental techniques [1,4,5,6,7,8,9,10]. Neutron diffraction experiments [4,6] recognized below this temperature the development of a commensurate antiferromagnetic (AFM) spin density wave (SDW) with a small moment in the FeAs plane. The reduction of the negative charge in the *RE*-O layer by F-substitution [1,11] or ipo-stoichiometric O-content [12][13] as well as the substitution of *RE* with both bi- and tetra-valent cationic species [14][15], suppresses the magnetic order while simultaneously superconductivity is set at lower *T*.

Since the first report in La($O_{1-x}F_x$)FeAs, an impressive effort was devoted to increase $T_c$ that, in fact, in less than two months was more than doubled, by substituting *RE* atoms with smaller ionic radius and tuning properly the doping [13,16]

Now that optimal $T_c$ seems reached in F-substituted SmFeAsO [17] with $T_c = 55$ K, a new scenario is opening which requires the optimization of the synthesis procedures, deeper characterization of products, in order to obtain controlled samples for more reliable investigations to face out the theoretical problem of the type of superconductivity, its relationship with the magnetic properties of the system, its similarity or not with the high $T_c$ superconductors.

In this paper we carried out a structural and microstructural investigation of SmFeAsO and SmFeAs($O_{0.93}F_{0.07}$) samples by X-ray diffraction, scanning and transmission electron microscopy as well as magnetic and transport properties characterization. The effect of sintering and the correlation between structural and physical properties is discussed.

## 2. Experimental

SmFeAsO and SmFeAs($O_{0.93}F_{0.07}$) were both prepared in two steps: 1) synthesis of SmAs starting from pure elements in an evacuated glass flask at a maximum temperature of 550°C; 2) synthesis of the oxy-pnictide reacting SmAs with stoichiometric amounts of Fe, $Fe_2O_3$ and $FeF_2$ at 1200°C for 24h in a form of a pellet in an evacuated quartz flask. In same cases, to improve the mechanical properties of the samples, the obtained oxy-pnictides were further pressed and sintered at 1300°C for 72h in an evacuated quartz flask. Phase identification was performed by X-ray powder diffraction, XRPD, (PHILIPS PW1830; Bragg-Brentano geometry; $CuK_{\alpha}$; range 20 – 110° $2q$ ; step 0.025° $2q$; sampling time 12 s); the crystal structures of both samples were refined in the space group $P4/nmm$ - 129 (origin choice 2) according to the Rietveld method using the program FullProf [18]. The diffraction lines were modelled by pseudo-Voigt functions and the background by a fifth-order polynomial. The following parameters were refined: the overall scale factor; the background (six parameters of the $5^{th}$ order polynomial); $2q$-Zero; the unit cell parameters; the specimen displacement; the half-width parameters; the peak shape; the reflection-profile asymmetry; the Lorentzian isotropic strain; the $z$ Wyckoff positions of Sm and O(F) (both located at the $2c$ site); the isotropic thermal parameters $B$. In order to obtain a good fit of the calculated diffraction pattern also the preferred orientation parameters were refined.

The microstructure of the samples was observed after metallographic preparation (embedding in epoxy resin and polishing with diamond paste) by means of a scanning electron microscope (SEM – LEICA Cambridge S360) equipped with an electron dispersive wavelength microprobe (EDS – OXFORD Link Pentafet). Transmission electron microscope (TEM) analysis was carried out by a JEOL Jem 2010 instrument operated at 200 kV. Each sample was powdered, dispersed in ethanol and the suspension finally deposited onto a carbon-coated copper grid. Selected area electron diffraction (SAED) patterns were collected along different zone axes on several particles; microprobe analysis was performed using an energy-dispersive X-ray spectrometer system (OXFORD Pentafet).

Magnetic measurements were performed by a SQUID magnetometer (MPMS by Quantum Design) in the temperature range 2 K–300 K. Resistivity measurements were made from 400 mK to 4 K in an adiabatic demagnetization cryostat and up to room temperature with a home made cryostat; in the whole temperature range a four probe technique was used. Magnetoresistivity measurements were performed in a PPMS Quantum Design system from 4 to 300 K in magnetic field up to 9 T.

## 3. Results and discussion
*3.1 XRPD and Rietveld refinement*
The XRPD pattern reveals the formation of both SmFeAsO and $SmFeAs(O_{0.93}F_{0.07})$; very few amounts of $Sm_2O_3$ can be detected as impurity. Rietveld refinement reveals a certain degree of preferred orientation along $00l$ in both samples and hence it is imperative to take into account this feature in order to obtain good $R$-factors values. The structural data of both samples obtained after refinement are reported in Table 1, whereas Fig. 1 shows the Rietveld plot of SmFeAsO, selected as representative. After F-substitution a notable contraction along the $c$-axis takes place, coupled with a very faint reduction of the $a$-axis. This feature is not related to a steric effect depending on F-substitution (ionic radii [19] in 4-fold coordination: $F^- = 1.31$ Å; $O^{2-} = 1.38$ Å). The comparison of the data reported in Table 1 reveals that the position of As is unaffected by substitution, as well as those of Fe and O being constrained by symmetry. Conversely Sm is shifted along the $c$-axis due to the decrease of the negative charge at the O site after F-substitution; hence the attraction between $Sm^{3+}$ and $As^{3-}$ is enhanced, as evidenced by the decreased length of the Sm-As bond (bridging the layers) coupled with a slight increase of the Sm-O one. As a result the two tetrahedral layers approach and the cell parameter $c$ decreases.

By applying the bond valence sum method (BVS; [20]) and using the inter-atomic distances obtained from the refined structural data, BVS calculations were performed at the $Sm^{3+}$ site: a good agreement (~3.15 v.u.) with the expected nominal valence is obtained considering $Sm^{3+}$ bonded to the neighbouring $O^{2-}$ and $As^{3-}$. This result indicates that the nominal valences of the species constituting the conducting layer are $As^{3-}$ (from the afore mentioned BVS calculations) and $Fe^{2+}$ (as required by charge balance). The Fe-As bond length does not change with F-substitution (Table 2).

*3.2 SEM observation*

Two specimens of the sample SmFeAsO were analysed: the former after synthesis (as prepared specimen) and the latter after sintering. The observation of the fractured specimen 1 shows that it is constituted of rectangular shaped tabular crystals of SmFeAsO with edge up to 20 μm (Fig. 2). The analysis of this sample after metallographic preparation reveals very few amounts of unreacted phases (iron arsenides) that are completely dissolved after sintering (specimen 2). In general sintering largely increases the density of the samples, up to ~85% (Fig. 3), but favours the formation of small particles of $Sm_2O_3$ (light particles in Figure 3, bottom). This feature reveals that at the sintering temperature the formation of $Sm_2O_3$ competes with the thermodynamic stability of the oxy-pnictide. A closer observation of the sintered samples reveals that also in this case they are constituted of randomly distributed lamellar crystals, thus explaining the preferred orientation along 00$l$ detected in the XRPD patterns; a contribution to this feature is in any case also originated by cleavage induced during the preparation of the powders for the analysis.

*3.3 TEM analysis*

Whatever the sample, the shape of the crystals and their cleavage favour the observation along [001], preventing the analysis parallel to the [100] zone axis, even though in some cases some crystal were successfully oriented along zone axes with $l \neq 0$. Fig. 4 shows a high resolution TEM (HRTEM) image of SmFeAsO viewed along [001] characterized by 2.78 Å periodicity consistent with (110) inter-planar spacing; the corresponding SAED patterns is reported in the inset. HRTEM images of SmFeAs($O_{0.93}F_{0.07}$) exhibits the same features of the pristine compound: Fig. 5 shows a crystal viewed along [201] exhibiting 3.90 Å and 2.89 Å periodicities, corresponding to (100) and (102) inter-planar spacing, respectively. Even though this is not the optimal orientation, an important conclusion can be pointed up: if extended planar defects along the *c* axis were present they should be evidenced both by the direct observation of the HRTEM images and by streaking in the corresponding SAED pattern, but these features are both lacking. In general no evidence for the occurrence of nanodomains, twinning and super-lattice reflections can be detected in both pristine and F-substituted samples. Conversely these kind of defects occurs and play a major role in the properties of HTSC-cuprates, where their formation is mainly related to oxygen stoichiometry [21,22,23]. It can be concluded that in our samples stoichiometric defects, if present, do not affect the microstructure at the atomic scale.

*3.4 Magnetic characterization*

Fig. 6 shows the molar susceptibility versus temperature for both specimens of SmFeAsO, as well as that of SmFeAs($O_{0.93}F_{0.07}$). The measurements were performed by applying a magnetic field of 3 T after a zero field cooling procedure. All measurements are affected by a temperature independent background magnetic signal, related to both the Pauli paramagnetic component and to a very small amount of ferromagnetic impurities. We estimated the ferromagnetic component by magnetization versus field measurements performed at $T = 5$ K (or 60 K for the superconducting Sm($O_{0.93}F_{0.07}$)FeAs sample) and $T = 300$ K. In these measurements (not shown here) a signal that linearly increases with the field is over-imposed to a ferromagnetic signal that is nearly the same at the two temperatures: it saturates to about the same value and at the same magnetic field (about 1 T), indicating that the ferromagnetic impurities have a transition temperature higher than 300 K. Only few parts over $10^4$ of Fe metal justify the observed behaviour. This temperature independent magnetic component has been subtracted to the measured curves obtaining the data of molar susceptibility shown in Figure 6.

The prominent characters of the molar susceptibility in both SmFeAsO specimens are the following: i) a flat maximum is present at $T \cong 140$ K, ii) a sharp peak is present $T = 6$ K and iii) a general decrease of magnetization when the temperature is increased is observed. The peak at 140 K marks the establishment of the AFM ordering in the Fe-As layers. The low temperature peak is due to the AFM ordering of Sm ions sublattice. This attribution is confirmed by specific heat measurements [24], where a low temperature peak ($T \cong 5$ K) is visible both in pristine and F-substituted SmFeAsO samples. We point out that AFM ordering of $Sm^{3+}$ sub-lattice is often found in superconducting [3] or not superconducting compounds [25]. Moreover also in Fe-pnictides system where $RE$ = Gd a low temperature peak is observed in magnetic measurements followed by a Curie–Weiss behaviour corresponding to the Gd paramagnetic behaviour [14]. Finally the decrease of magnetization with temperature is mainly due to the paramagnetic behaviour of the Sm ions that, differently than in the case of Gd or other $RE$, does not follow a Curie-Weiss behaviour [26]. Above $T = 140$ K also a paramagnetic behaviour of Fe is expected, which should be small in the light of the value of 0.2÷0.3 $\mu_B$ attributed to Fe magnetic moment in the ordered state [5]. The effect of sintering on the magnetic properties of SmFeAsO is not only evidenced by the shift of the $\chi$ versus $T$ curves, but also by the different concavity observed in various parts of these curves. We cannot exclude that a small quantity of spurious magnetic phases, that are present in the samples as evidenced by SEM analysis, (mainly $Sm_2O_3$ in undoped and SmOF in substituted samples, both exhibiting paramagnetic behaviour above $T = 2$ K) contribute to the magnetic signal producing the observed differences.

In SmFeAs($O_{0.93}F_{0.07}$) the peak at $T \cong 140$ K disappears, according to what generally observed and the superconducting transition appears, whose onset is about 32 K at 3 T. Moreover the overall magnetic moment is increased respect to the SmFeAsO samples. This effect has been observed in LaFeAsO samples substituted with various amount of F [11]. It is not yet clear if this increase is related to the presence of some magnetic impurities and/or to a novel magnetic behaviour arising in the Fe-As layers.

*3.5 Resistivity*

Differently from magnetization measurements, normal state resistivity are unaffected by insulating spurious phases. Figure 7 shows resistivity versus temperature measurements of SmFeAsO, before and after sintering, as well as that of SmFeAs($O_{0.93}F_{0.07}$). SmFeAsO specimens exhibit a pronounced anomaly around $T\sim 140$ K, i.e., at the same temperature of the anomaly shown by magnetization measurements, then resistivity decreases monotonically with decreasing temperature. SmFeAs($O_{0.93}F_{0.07}$) is characterized by a nearly linear decrease of resistivity with temperature, with the onset of superconductivity occuring at 36 K.

The as prepared SmFeAsO specimen was measured down to 400mK (see the inset in figure 7). Saturation is not achieved neither at low temperature; in addition below 6 K a drop is observed, which could be related to the suppression of magnetic scattering of Sm moments ordering below this temperature (see fig.6).

As discussed previously, the specimens of SmFeAsO underwent different thermal treatments and this is also well emphasized by resistivity measurements. The sintered specimen is characterized by resistivity values nearly three times lower than the as prepared one. As shown by SEM analyses, sintering process produces a phase densification and this improve connections among grains which well explains the reduced resistivity. As a first approximation poor connectivity increases the measured resistivity since the effective sample cross section is reduced; thus resistivity data have to be normalized to compare the intragrain resistivity of the two specimens. Figure 8 shows $\rho(T)/\rho_{peak}$ versus temperature of SmFeAsO specimens, where $\rho_{peak}$ are the resistivity values at $T_{peak} \sim 147$ K. This plot emphasizes the high degree of crystallinity of both specimens, whose resistivity drop from $T_{peak}$ to 4 K is more than doubled. This confirms the TEM analysis which evidenced the lacking of structural defects. The as prepared specimen shows a lower residual resistivity which is indicative of a purer phase, (see SEM analysis).

In order to analyse the temperature dependence of resistivity the residual resistivity $\rho_0$ should be subtracted. The inset of Figure 8 shows that residual resistivity is not yet reached at the minimum temperature explored in the measurements, *i.e.* 400 mK. Thus for both specimens we assume, in some sense arbitrarily, $\rho_0=0.95\rho(5$ K$)$: in the inset of Figure 8 we plot $\rho(T)-\rho_0$ versus $T$ from 5 K to room temperature. Both curves present similar behaviour and the main features are not affected by a different choice of $\rho_0$: below 140 K the resistivity curves decrease roughly as a power law $T^n$ with the coefficient $n$ close to 2. This behaviour is lost below 20 K where the curves flatten. Within the SDW framework the suppression of resistivity is due to the combined effects of carrier condensation and increased mobility of electrons which do not participate in the SDW. This produces the power law behaviour, while the abrupt flattening of resistivity below 20 K could depend on the increasing scattering from spin fluctuations related with Sm ordering below 6 K.

Magnetoresistivity measurements were performed on the as prepared SmFeAsO specimen from 0 to 9 T in order to evaluate the upper critical field, $B_{c2}$. Figure 9 shows magnetoresistivity measurements at fixed magnetic field varying the temperature. The transition is rather enlarged also in zero field indicating a not homogeneous fluorine distribution. $B_{c2}$, evaluated at the 90% of the resistive transition and at the maximum of $d\rho/dT$ is shown in the inset. Both curves rise steeply showing an upturn curvature. It was proposed [27] that this curvature is related to multiband nature of superconductivity in this class of compounds; in our sample distribution of grains with different $T_c$ could affect the observed behaviour. A unique extrapolation of $B_{c2}(T=0)$ can not be obtained by the curvature. If we consider $B_{c2}$ at the 90% of the resistive transition and we evaluate an average slope $dB_{c2}/dT \sim 4$ T/K, the corresponding $B_{c2}(T = 0)$ value derived from the Werthamer-Helfand-Hohenberg (WHH) formula is $0.693 T_c | dB_{c2}/dT |_{T_c} \approx 100$ T. Similar values are reported for the same compound in [28]. Considering that the WHH evaluation underestimates $B_{c2}$ in two band superconductors [29], huge upper critical fields are expected in these compounds. More certain evaluations will require magnetoresistivity measurements at much higher applied magnetic field.

**4. Conclusions**

SmFeAsO and SmFeAs(O$_{0.93}$F$_{0.07}$) were synthesized and their structures were refined by applying the Rietveld method; the effect of sintering on microstructure was investigated by means of both SEM and TEM analysis. Sintering largely increases (up to 85%) the density of the samples improving mechanical properties; on the other hand it induces a competition between the thermodynamic stability of the oxy-pnictide and Sm$_2$O$_3$. Indeed small particles of Sm$_2$O$_3$ are emphasized by SEM analysis in the sintered sample whereas are not present in the as prepared one. TEM analysis shows no evidence for the occurrence of nanodomains, twinning and super-lattice reflections. All this frame is well confirmed by magnetization and resistivity measurements. In the SmFeAsO samples magnetization measurements exhibit a flat maximum around 140 K which marks the establishment of the AFM ordering in the Fe-As layers, and a peak around 6 K due to the AFM ordering of Sm ions sublattice. In SmFeAs(O$_{0.93}$F$_{0.07}$) the feature at 140 K is absent, while superconductivity appears below 32 K.

Resistivity measurements of pristine specimens present the same features observed by magnetization. In fact, not only a sharp maximum around 140 K is well observable, but measurements down to 400 mK are able to discern a drop of resistivity below 6 K. This is a clear indication that carriers localized in FeAs layer interact with Sm ions out of plane. Effect of sintering are well emphasized by resistivity measurements: the sintered sample presents much lower resistance due to the increased connection between grains, but slightly higher residual resistivity due to defects induced by sintering. In both cases the sharp drop of resistivity below 140 K is indicative of high purity phase in agreement with TEM analysis.

Finally, resistivity of SmFeAs(O$_{0.93}$F$_{0.07}$) decreases nearly linear with temperature and shows the onset of superconductivity below 36 K. Magnetoresistivity measurements performed up to 9 T suggest the occurrence of upper critical field of about 100 K; only measurement performed in high magnetic field will confirm these preliminary results.

Table 1: Strutural data of SmFeAsO and SmFeAs($O_{0.93}F_{0.07}$) obtained after Rietveld refinement using XRPD data ($Z = 2$).

|  |  |  | **SmFeAsO** |  |  | **SmFeAs($O_{0.93}F_{0.07}$)** |  |  |
|---|---|---|---|---|---|---|---|---|
| Cell edge | $a$ (Å) | | 3.9391(2) | | | 3.9344(2) | | |
|  | $c$ (Å) | | 8.4970(4) | | | 8.4817(5) | | |
|  |  |  | $x$ | $y$ | $z$ | $x$ | $y$ | $z$ |
| Wyckoff position | 2c | Sm | ¼ | ¼ | 0.1368(2) | ¼ | ¼ | 0.1397(2) |
|  | 2a | O/F | ¾ | ¼ | 0 | ¾ | ¼ | 0 |
|  | 2b | Fe | ¾ | ¼ | ½ | ¾ | ¼ | ½ |
|  | 2c | As | ¼ | ¼ | 0.6609(4) | ¼ | ¼ | 0.6611(5) |
| R-Bragg (%) | | | 4.92 | | | 5.05 | | |
| Rf-factor (%) | | | 4.23 | | | 4.29 | | |

Table 2: Comparison among selected bond lengths in SmFeAsO and SmFeAs($O_{0.93}F_{0.07}$).

|  | **SmFeAsO** | **SmFeAs($O_{0.93}F_{0.07}$)** |
|---|---|---|
| Sm-O × 4 (Å) | 2.287(1) | 2.296(1) |
| Sm-As × 4 (Å) | 3.273(2) | 3.255(2) |
| Fe-As × 4 (Å) | 2.397(2) | 2.395(2) |

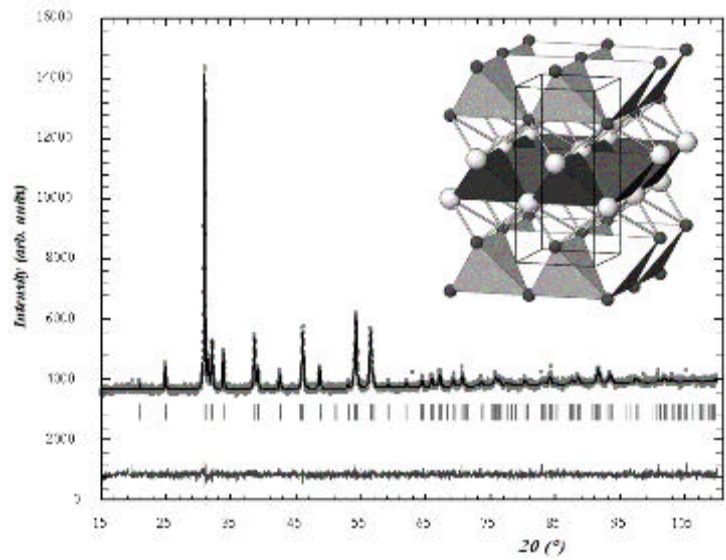

Figure 1: Rietveld refinement plot for SmFeAsO; the inset shows a clinographic view of the crystal structure: dark and light tetrahedra are centred by Sm and Fe, respectively; O and As are represented by dark and light spheres, respectively.

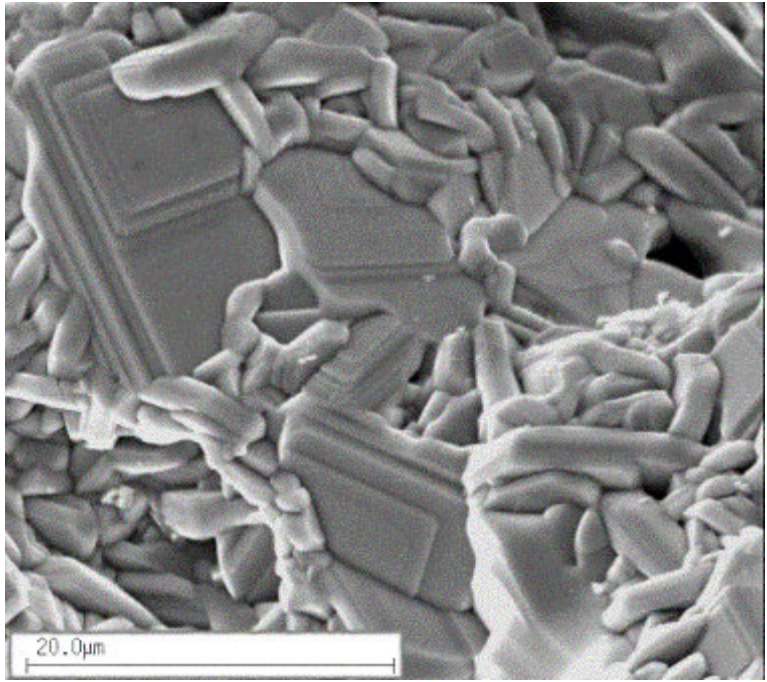

Figure 2: SEM image (secondary electrons; fractured sample) showing lamellar crystals of SmFeAsO (as prepared sample).

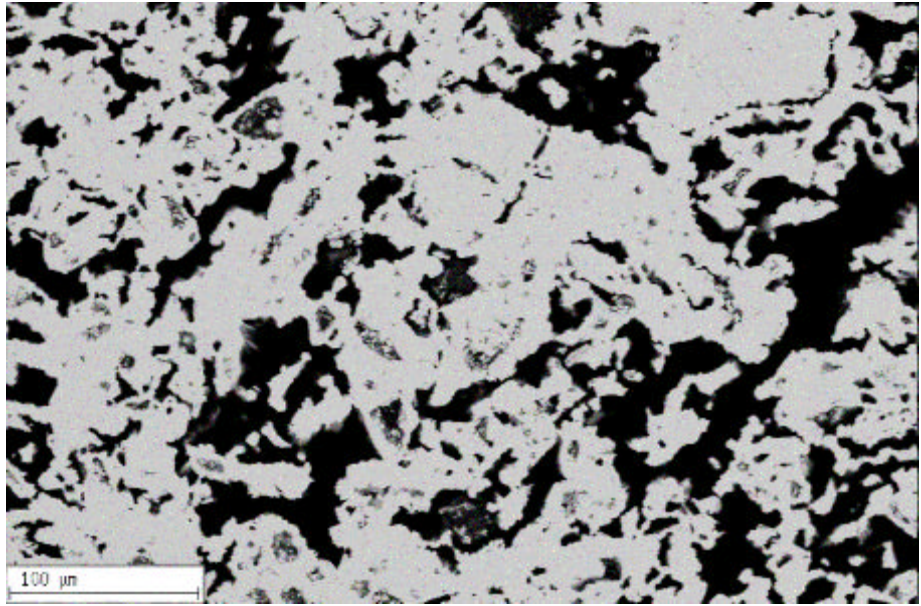

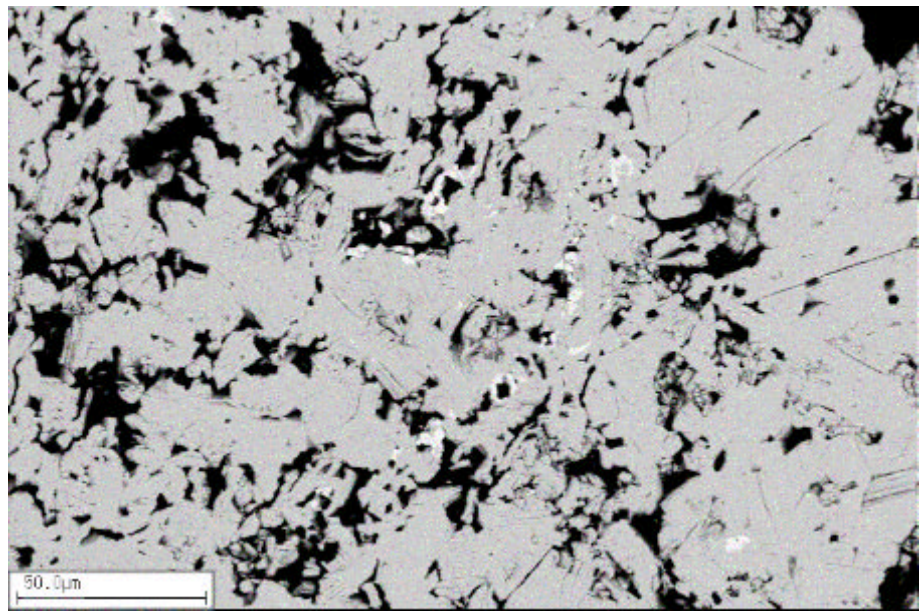

Figure 3: Comparison of the microstructures of SmFeAsO in the as prepared sample (top) and after sintering (bottom); dark zones are voids filled up by epoxy resin.

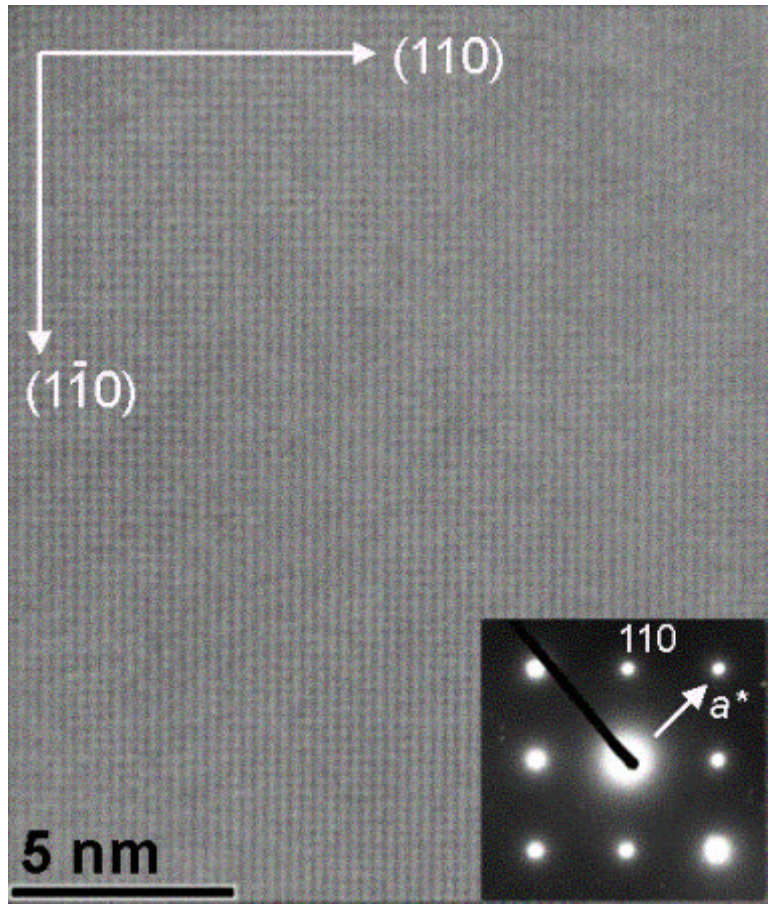

Figure 4: HRTEM image of SmFeAsO viewed along [001] showing 110 periodicity; the inset is the corresponding indexed SAED pattern.

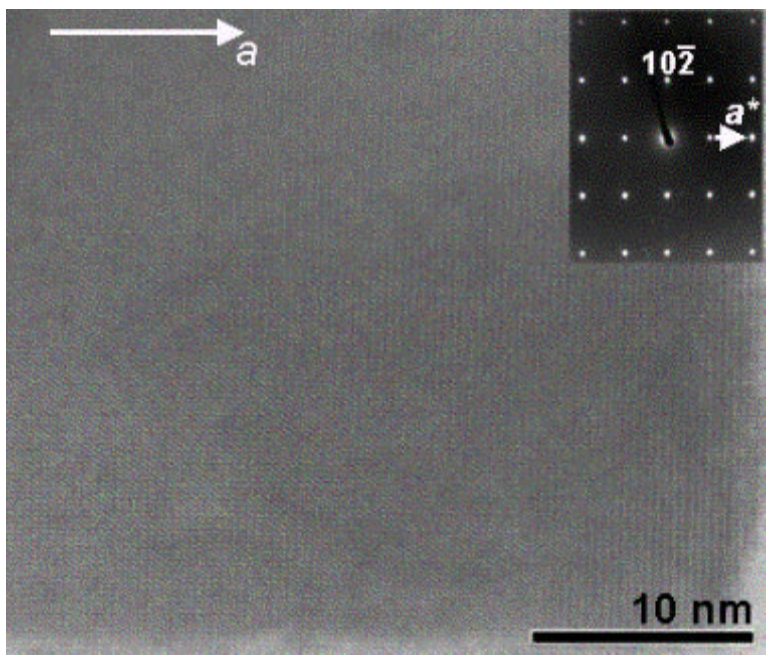

Figure 5: HRTEM image of SmFeAs($O_{0.93}F_{0.07}$) viewed along [201] showing 100 and 102 periodicity; the inset is the corresponding indexed SAED pattern.

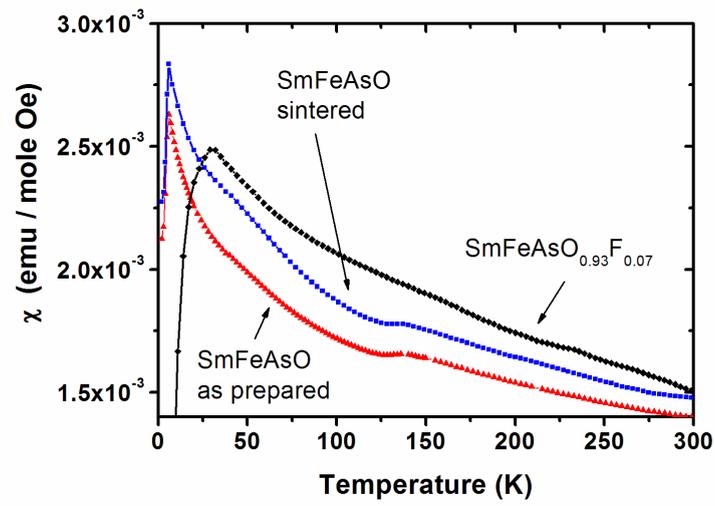

Figure 6: Molar susceptibility versus temperature for both pristine and doped samples. A background temperature independent magnetic signal has been subtracted to each curve as explained in the text

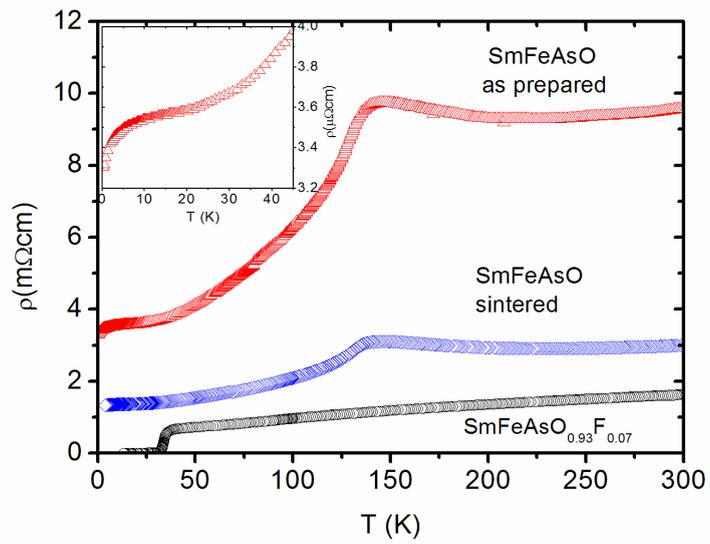

Figure 7: Resistivity versus temperature of SmFeAsO, before and after sintering, as well as that of SmFeAs($O_{0.93}F_{0.07}$). Inset: magnification of the low temperature region of SmFeAsO, before sintering,

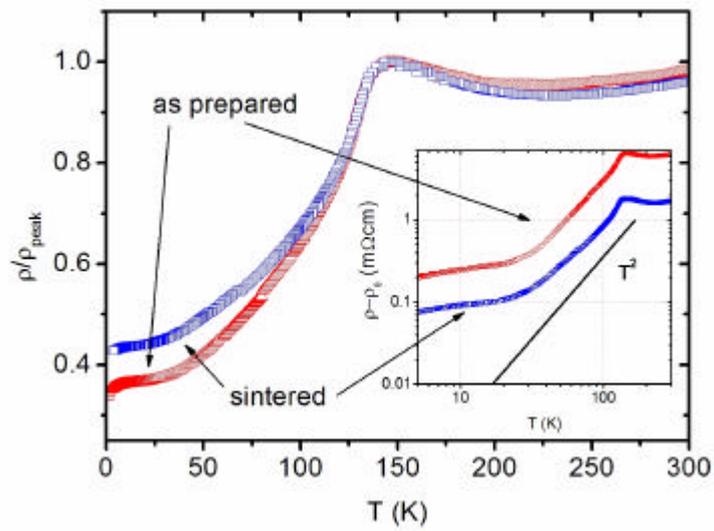

Figure 8: $\rho(T)/\rho_{peak}$ versus temperature of SmFeAsO specimens where $\rho_{peak}$ are the resistivity values in correspondence of the peak (for both samples at 145-149 K). Inset: $\rho(T)-\rho_0$ versus temperature where $\rho_0=0.95\rho(5\ K)$.

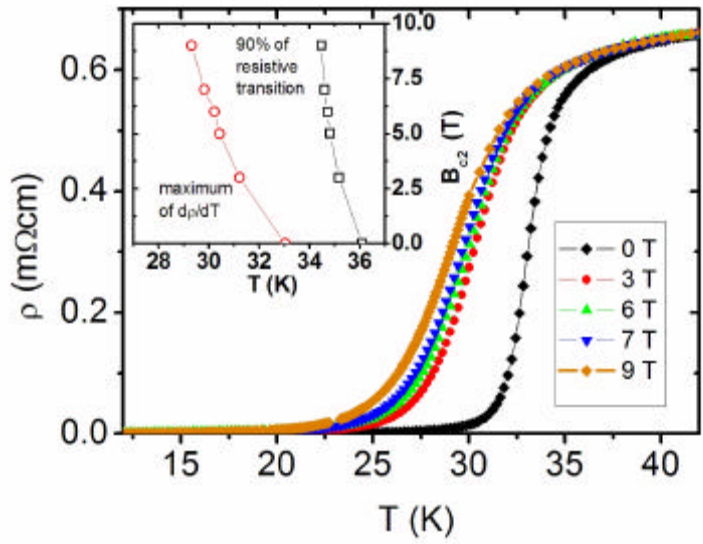

Figure 9: Magnetoresistivity measurements at fixed magnetic field versus temperature for SmFeAs($O_{0.93}F_{0.07}$). Inset: $B_{c2}$ vs temperature evaluated at the 90% of the resistive transition and at the maximum of d$\rho$/dT.